# Electric-Magnetic-Switchable Free-Space Skyrmions in Toroidal Light Pulses via a Nonlinear Metasurface

Li Niu[1,#], Xi Feng[1,#], Xueqian Zhang[1,*], Wangke Yu[2], Qingwei Wang[1], Yuanhao Lang[1], Quan Xu[1], Xieyu Chen[1], Jiajun Ma[1], Haidi Qiu[1], Yijie Shen[2,3,*], Weili Zhang[4,*], Jiaguang Han[1,5,*]

[1] Center for Terahertz Waves, State Key Laboratory of Precision Measurement Technology and Instruments, Tianjin University, Tianjin 300072, China
[2] Centre for Disruptive Photonic Technologies, School of Physical and Mathematical Sciences, Nanyang Technological University, Singapore 637371, Singapore.
[3] School of Electrical and Electronic Engineering, Nanyang Technological University, Singapore 639798, Singapore.
[4] School of Electrical and Computer Engineering, Oklahoma State University, Stillwater, Oklahoma 74078, USA
[5] Guangxi Key Laboratory of Optoelectronic Information Processing, School of Optoelectronic Engineering, Guilin University of Electronic Technology, Guilin 541004, China

[#]These authors contributed equally to this work
[*]Corresponding authors. Email: alearn1988@tju.edu.cn, yijie.shen@ntu.edu.sg, weili.zhang@okstate.edu, jiaghan@tju.edu.cn

## Abstract

Recent advances reveal that light propagation in free space supports many exotic topological textures, such as skyrmions. Their unique space-time topologies make them promising candidates as next-generation robust information carriers. Hence, the ability of switching different texture modes is highly demanded to serve as a manner of data transfer. However, previous studies focus on generation of one specific mode, lacking integrated devices with externally variable and stable mode generation capability. Here, we experimentally demonstrate the first realization of switchable skyrmions between electric and magnetic modes in toroidal light pulses using a nonlinear metasurface platform in terms of broadband terahertz generation driven by vectorial pulse. The spatial and temporal evolutions of them are also clearly observed. Our work establishes a new paradigm for manipulating and switching topologically structured light.

# Introduction

Topological textures, including skyrmions[1,2], merons[3,4], and hopfions[5,6], have emerged as a research frontier due to their intrinsic stability, controllable dynamics, and robustness against perturbations governed by conservation laws, making them ideal candidates for a wide range of information storage technologies[7]. Originally discovered in magnetism, these concepts have recently drawn huge interest in electromagnetic realm, especially in free-space propagating topological textures, which enable long range dynamic transfer of topologies[8]. By exploiting optical spin angular momentum[9,10], vectorial electric fields[11,12], polarization Stokes vector[13,14], energy flow vector[15], momentum vector[16], and pseudo-spin vectors[17], researchers have demonstrated a rich variety of topological textures that further advances the structured-light control. In particular, optical skyrmions—topologically protected, soliton-like field structures—exhibit remarkable stability, making them promising candidates for robust information encoding and processing[1]. However, despite their significance, previous studies have predominantly focused on generating skyrmions in evanescent waves or static vector beams, where the spatial extent and tunability are fundamentally constrained[9,11,18]. Extending skyrmion generation and control to free space is therefore an important step toward exploring their broader application potential[8,19,20], especially in long-range information transfer. To fully harness their potential, it is therefore crucial to establish a platform that can simultaneously generate and dynamically manipulate the skyrmion textures.

An effective solution is offered by toroidal light pulses (TLPs)—space-time non-separable, single-cycle electromagnetic pulses with toroidal topology[21-26]. Unlike conventional transverse waves, TLPs possess strong longitudinal field components and 3D vectorial structure, making them ideal hosts for complex topological textures. Their nonparaxial nature and intrinsic toroidal geometry allow skyrmions to propagate in free space as localized energy packets with preserved topological identity[23-26]. However, their experimental generation and characterization have long been hindered by their intricate spatiotemporal structure, ultrashort duration, and topological constraints.

Recent advances have overcome these challenges using metasurface method in both the optical and terahertz (THz) regimes[21]. Parallel breakthroughs happen in semiconductor platforms, which demonstrate the generation of THz TLPs via quantum interference driven by tailored two-color laser fields[22]. Notably, free-space skyrmionic textures have also been realized using specially designed axial horn antennas in the microwave regime[26]. Despite these advancements, previous studies have primarily focused on individual topological structures, such as transverse magnetic (TM) or transverse electric (TE) TLPs, corresponding to electric skyrmion (E-skyrmion) or magnetic skyrmion (H-skyrmion), without addressing the dynamic control or switching of these states within an integrated system.

In this work, we present the first experimental demonstration of on-demand switchable generation of E-skyrmion and H-skyrmion in free-space propagating single-cycle THz TLPs, as shown in Fig. 1. This achievement is enabled by a tailored spatially multiplexed nonlinear metasurface that is selectively excited by azimuthally or radially structured near infrared (NIR) pump beams, thereby establishing real-time switching between TM and TE TLPs topologies. Through spatiotemporal characterization, we reveal propagation-dependent evolution of E-skyrmion and H-skyrmion in free space, including their deterministic polarity reversal feature. Besides, the vectorial pumps are also promising by their outstanding property in maintaining polarization and intensity structures upon propagation, making the whole information transfer process from optical to terahertz stable to external disturbance. This breakthrough establishes a foundation for topology-encoded optical communication systems and integrated platforms for both quantum and classical information applications.

## Results

TLPs are also free-space solutions to the homogeneous Maxwell's equations that exhibit space-time non-separability and a unique configuration of electric and magnetic fields[27-29]. These pulses are characterized by their non-transverse electromagnetic components, which are tightly coupled with their space-time structure, as well as by their single-cycle nature. TLPs can exist in both TM and TE modes, with each mode displaying distinct field characteristics. The electric and magnetic fields of TM and TE TLPs are described as follows[27]:

$$E_z^{TM} = \sqrt{\frac{\mu_0}{\varepsilon_0}} H_z^{TE} = -4f_0 \sqrt{\frac{\mu_0}{\varepsilon_0}} \frac{r^2 - (q_1 + i\tau)(q_2 - i\sigma)}{[r^2 + (q_1 + i\tau)(q_2 - i\sigma)]^3}, \quad (1)$$

$$E_r^{TM} = \sqrt{\frac{\mu_0}{\varepsilon_0}} H_r^{TE} = 4if_0 \sqrt{\frac{\mu_0}{\varepsilon_0}} \frac{r(q_2 - q_1 - 2iz)}{[r^2 + (q_1 + i\tau)(q_2 - i\sigma)]^3}, \quad (2)$$

$$H_\theta^{TM} = -\sqrt{\frac{\varepsilon_0}{\mu_0}} E_\theta^{TE} = 4if_0 \frac{r(q_1 + q_2 - 2ict)}{[r^2 + (q_1 + i\tau)(q_2 - i\tau)]^3}, \quad (3)$$

where $\sigma = z + ct$, $\tau = z - ct$, $(r, z)$ represents the cylindrical coordinate, $\mu_0$ and $\varepsilon_0$ represent the vacuum permeability and permittivity, $t$ is time, $f_0$ is a normalization constant, $q_1$ and $q_2$ denote the effective wavelength and Rayleigh range, respectively. Figure 1 visually presents the spatial distributions of these fields in both TM and TE configurations, with the yellow and blue regions representing the magnetic and electric fields, respectively. In the TM mode, the magnetic field exhibits an azimuthal polarization, while the electric field is predominantly radial and longitudinal. In contrast, the TE mode features an azimuthally polarized electric field, with the magnetic field primarily oriented in the radial and longitudinal directions. Here, this mode switching is enabled by a simple polarization control of the incident pump beam. Initially, a *y*-polarized pump (*y*-pump) is converted into an azimuthally polarized pump beam (A-pump) using a vortex half-wave retarder (VHR). When this beam excites the nonlinear metasurface, it emits a TM-polarized TLP that carries a propagating E-skyrmion. By rotating the half-wave plate (HWP), the incidence is switched to *x*-polarized pump (*x*-pump), which is then transformed into a radially polarized pump beam (R-pump) by the same VHR. This beam drives the nonlinear metasurface to emit a TE-polarized TLP that carries a propagating H-skyrmion.

**Design of the nonlinear metasurface**

To generate TLPs, we employed nonlinear metasurfaces constructed from gold splitting-ring resonators (SRR) (see Fig. 1), which offer a flexible mechanism for controlling the spatiotemporal polarization properties of the generated THz TLPs[30-32]. Figure 2a illustrates the unit cell of the nonlinear metasurface, comprising a SRR patterned on an 8-nm-thick ITO layer deposited on a fused quartz substrate. The geometric parameters of the structure are $l_x$ = 220 nm, $l_y$ = 212 nm, $l$ = 110 nm, $w$ = 79 nm, $t_{gold}$ = 40 nm, $t_{ITO}$ = 8 nm, $t_{glass}$ = 0.7 mm, and $P$ = 420 nm, respectively. The simulated polarization-dependent transmission spectra are shown in Fig. 2b. Under *x*-polarized illumination, the structure exhibits a magnetic dipole resonance centered at a wavelength of 1350 nm. When the pump beam illuminates the SRR, it acts as a nonlinear source, radiating THz waves. Specifically, as shown in the generation mechanism in Fig. 2c[30], when an *x*-pump illuminates, the electric field drives charge movement along the SRR arms and base, generating a loop linear current (red) that forms a magnetic dipole response. This, in turn, results in nonlinear currents (blue) flowing in the same direction along both arms since they are parallel or anti-parallel to the increasing or decreasing charge movement, leading to in-phase *y*-polarized radiations towards the far field. In contrast, when an *y*-pump illuminates, the electric field drives charge movement only in the two arms symmetrically. In this case, the nonlinear currents flow in opposite directions along each arm, causing zero far-field radiation by destructive interference.

Leveraging the above polarization-selective THz generation characteristics of the SRR, precise control over charge flow can be achieved by arranging the unit cells in tailored spatial configurations. This capability lays the foundation for actively manipulating vectorial currents at nanometer spatial scales and femtosecond timescales, providing a flexible method for tailoring the emitted THz field[33]. Figures 2d and 2e illustrate the SRR arrangements for generating TM and TE TLPs, respectively. For generating TM TLP, SRRs are arranged concentrically with arms aligned radially (Pattern 1). An A-pump NIR beam induces nonlinear currents along the radial SRR arms, radiating a TM TLP with a radially distributed THz electric field. For generating TE TLP, SRR arms are arranged azimuthally (Pattern 2). A R-pump NIR beam induces nonlinear currents in the azimuthal direction, resulting in a TE TLP with an azimuthally distributed THz electric field. To realize the switchable TM and TE THz TLP emission within a single device, spatial multiplexing is employed to integrate Pattern 1 and

Pattern 2, as shown in Fig. 2f. To validate the design, we fabricated a 1 mm × 1 mm sample (see Supplementary Note 1 for fabrication details). The scanning electron microscope (SEM) image of its central region is shown in Fig. 2g, where the radial and angular arrangements of the SRRs can be directly distinguished by the light blue and light red (false color) regions, respectively.

**Switching between TM and TE TLP generations**

To check whether the propagation dynamics of the emitted THz beams are in the desired spatiotemporal forms, we established a THz time-domain spectroscopy system to measure the space-time distribution of the generated free-space THz field from the fabricated sample (See Supplementary Note 2 for experimental setup details). Figure 3a illustrates the schematic of the space-time evolution of the second-type transverse TLP propagating along the +$z$ direction that fits our case[21]. It can be seen that such a TLP has 1½ cycles at the focal point ($z$ = 0 mm), while 1 cycle at $z = \pm q_2/2$. In addition to the number of cycles, distinct focusing and diverging wavefronts are observed at $z <$ 0 mm and $z >$ 0 mm, respectively. This reshaping of the field distribution is attributed to the propagation-dependent Gouy phase term[34].

As a first step, we measure the space-time and space-spectral electric field distributions of the emitted THz pulse. Figures 3b to 3f show the measured 3D space-time amplitude |$E$| distributions of the transverse electric fields of the TM TLP at five distinct $z$-positions from $z = -2$ to $z = 2$ mm with an interval of 1 mm under the A-pump (See Supplementary Note 3 for experimental details). The corresponding measured 3D space-time electric field distributions of $E_x$ and $E_y$ are illustrated in Supplementary Fig. S4. It is evident that all THz transverse electric fields exhibit characteristic hollow distributions. At $z$ = 0 mm, the measured pulse exhibits 1½ cycles with a time-symmetrical distribution and an almost plane wavefront (see Fig. 3d). At $z$ = +1 mm (−1 mm), the pulse still shows a 1½ cycles but its temporal profile becomes asymmetrical and exhibits a diverging (focusing) wavefront (−1 mm), see Fig. 3e (Fig. 3c). At $z$ = +2 mm (−2 mm), the pulse nearly reduces to a single cycle, which exhibits greater time asymmetry and a more pronounced diverging (focusing) wavefront, see

Fig. 3f (Fig. 3b). Moreover, the overall spatial size of the pulse clearly reaches minimum at $z = 0$ mm, indicating the focal plane.

For clearer visualization of the spatiotemporal distribution of different electric field components, cross-sectional field distribution on the $y = 0$ plane is extracted. The calculated and measured normalized 2D space-time distributions of the transverse electric field ($E_x$) of the pulse on the $y = 0$ mm plane are illustrated in Figs. 3g and 3h, respectively. It is seen that the $E_x$ field exhibits clear out-of-phase behavior between the regions of $x > 0$ mm and $x < 0$ mm, while the pulse duration corresponds to 1½ cycles. Notably, the calculated results are derived from the TM TLP solutions of Maxwell's equations, with $q_1$ set to 70 μm and $q_2$ to $60q_1$. It is seen that the measured $E_x$ field agrees well with the theoretical prediction. Besides, the $E_x$ and $E_y$ distributions of the TM TLP on the $x = 0$ and $y = 0$ planes are also found to be consistent with the theory (see Supplementary Fig. S5). To further verify the space-time non-separability, we extracted the average intensity distribution at different radii and frequencies from the Fourier transform results shown in Fig. 3d (see Supplementary Note 4 for details). This process yields a 2D THz intensity distribution, as shown in Fig. 3j. The corresponding calculated distribution, obtained using the optimally fitted parameters of $q_1$ and $q_2$, is illustrated in Fig. 3i. As their insets, both the blue lines and hollow blue dots represent the positions in the intensity maxima of the calculated and measured THz pulses at each frequency, respectively, which are in close agreement.

To quantify the similarity between the generated pulse and the ideally calculated TLP, we employ a state tomography-based method recently introduced for characterizing the space-spectrum non-separability in electromagnetic waves[28]. For an ideal TLP, the tomography matrix is perfectly diagonal, indicating that the spatial and spectral states are perfectly overlapping and non-separable, resulting in a fidelity of F = 1.00 (see Fig. 3k). For the measured THz pulse, the tomography matrix is nearly diagonal, yielding a high fidelity of F = 0.86 (see Fig. 3l). All the above results well indicate that the generated THz pulse reproduce the main features of the ideal THz TLP, showing the effectiveness of our nonlinear current control method using nonlinear metasurface in generating real THz TLP.

We further measured the corresponding results of the sample under R-pump at $z = 0$ mm, and compared them with calculations performed using the optimally fitted fitting parameters of TE TLP, where $q_1$ is set to 65 μm and $q_2$ to $60q_1$ (see Figs. 3m-3r and Supplementary Fig. S7). It is clearly seen that they are also agreeing well with each other. The space-spectral non-separability fidelity for the measured THz pulse is found to be F = 0.99. All the observed features in both the space-time and space-spectral distributions are also consistent with the space-time non-separable nature of the TE TLP. Notably, disruptions in the space-spectral distribution around 1.2 THz are attributed to water vapor absorption. This limitation can be mitigated by employing a desiccation chamber for the THz path. Overall, this device clearly demonstrates the potential for generating high-quality switchable TM and TE THz TLPs by simply altering the vectorial features of the pump.

**Switchable E- and H-skyrmion generations**

E-skyrmion and H-skyrmion are accompanying products of TM and TE TLPs, which guarantee the stability of the TLP propagations. Their switchability naturally follows that of the TLP modes. To experimentally observe their propagation property in space-time, it is essential to capture the spatiotemporal distributions of both the electric field components ($E_x$, $E_y$, $E_z$) and the magnetic field components ($H_x$, $H_y$, $H_z$). Such comprehensive measurements allow for the reconstruction of the full spatiotemporal vector field, which is crucial for visualizing and analyzing skyrmionic textures. To characterize the E-skyrmion in the TM TLP, we extracted the measured spatiotemporal distributions of $E_x$ and $E_y$ at $z = 0$. The $E_z$ component can then be derived using a transformation based on the Gauss's law[21]:

$$E_z(x,y,z) = -\int_\alpha^z (\frac{\partial E_x(x,y,z)}{\partial x} + \frac{\partial E_y(x,y,z)}{\partial y})dz , \qquad (4)$$

where $\alpha$ is chosen as a point at which the field is zero. Figure 4a shows the calculated $E_z$ component propagating over time on the $y = 0$ plane, where $E_z$ transitions from negative to positive values. Further analysis of the electric field distributions of different polarized components in the $xy$ plane at two distinct time points, $t_1$ and $t_2$, are shown in Figs. 4d and 4e. It is observed that the $E_x$ and $E_y$ components at $t_2$ almost recover the same feature of those at $t_1$, whereas the $E_z$ component undergoes a reversal in its distribution. The ideal electric field distributions, calculated using Eqs. 1 and 2, are shown in the insets, and they show good agreements with the experimental results. By analyzing the $E_x$, $E_y$, and $E_z$ components, we can reconstruct the 3D vector configuration of the skyrmion electric field, as shown in Figs. 4b and 4c. The skyrmionic textures at times $t_1$ and $t_2$ evolve during propagation while maintaining the Néel-type helicity in different transverse planes. At $t_1$, the electric vector of the skyrmion points downward at the center and gradually points upward at the periphery, whereas at $t_2$, the corresponding directions reverse. The topological characteristics of a skyrmionic configuration can be quantified by the skyrmion number $N_{sk}$[1] (See Supplementary Note 5 for details), which measures the similarity to an ideal skyrmion, and serves as an indicator of the beam quality. The calculated skyrmion numbers of −0.990 at $t_1$ and 0.992 at $t_2$ are close to ±1, satisfying the condition for an almost integer skyrmion number. This result confirms the presence of well-defined skyrmionic textures in the generated TM TLP.

When switched to the TE TLP, the nonlinear metasurface emits azimuthal electric field ($E_\theta$), whereas the magnetic field comprises both radial and longitudinal components ($H_r$ and $H_z$), thereby forming a magnetic skyrmion. To validate the experimentally observed H-skyrmion, the distributions of the magnetic field components $H_x$, $H_y$, and $H_z$ needs to be obtained. These components can be calculated from the measured electric field using the Faraday's law in free space. The calculated $H_z$ field propagating over time using the measured 2D spatiotemporal distributions of $E_x$ and $E_y$ from the TE TLP is shown in Fig. 4f. The $H_z$ component also reverses its direction over time. At times $t_3$ and $t_4$, the calculated distributions of $H_x$, $H_y$, and $H_z$ in the $xy$-plane are presented in Figs. 4i and 4j. Similarly to the E-skyrmion, $H_x$ and $H_y$ are nearly the same while the direction of $H_z$ flips with time. The inset in the lower-right corner of Figs. 4i and 4j shows the ideal distributions of the magnetic field components ($H_x$, $H_y$, $H_z$) calculated from Eqs. 1and 2, where excellent agreement with the measured results are observed. From these magnetic field components, the H-skyrmion texture is reconstructed, as shown in Figs. 4g and 4h. At $t_3$, $H_z$ reverses its direction featured by transitioning from pointing downward at the center to upward away from the center. At $t_4$, the magnetic vector changes direction from pointing upward at the center to downward away from the center. Since $H_x$ and $H_y$ remain, the H-skyrmion remains its Néel-type configuration. The calculated skyrmion number are –0.991 at $t_3$ and 0.994 at $t_4$, which are also close to ±1, confirming the presence of good H-skyrmionic textures in the generated TE TLP. All the above results validate the effectiveness of our method in generating switchable E-skyrmion and H-skyrmion very well.

## Discussion

The experimental realization of switchable E-skyrmion and H-skyrmion in free-space propagating TLPs marks a significant breakthrough in topological photonics. The intrinsic topological stability of these skyrmions may help maintain robust information transfer under scattering media and atmospheric turbulence, thereby overcoming a major limitation of conventional free-space optical links[1,8,19]. This behavior parallels the well-established robustness of skyrmions in magnetism, which exhibit strong resistance to thermal fluctuations and external disturbances, making them ideal for information storage[7]. These advantages offer strong potential for information transfer in noisy environments where traditional structured beams, such as Gaussian or vortex beams, are prone to distortion and performance degradation[35,36].

Beyond their stability, the switching ability between E-skyrmion and H-skyrmion introduces an additional degree of freedom for encoding information, enabling topologically multiplexed communication[37]. Unlike conventional polarization or mode-division multiplexing, this approach can enhance communication capacity while maintaining high signal integrity. Furthermore, the interplay between TLPs and skyrmions opens avenues for advanced encryption techniques in secure optical communication by utilizing their unique spatial and topological characteristics.

A key aspect of our work lies in the potential versatility of nonlinear metasurfaces as a design platform. By precisely engineering their properties, they may enable the generation and modulation of a variety of topological states, with the possibility of extending to higher-dimensional structures[1,23]. This could offer new opportunities for controlling skyrmion configurations and studying light-matter interactions across distinct topological regimes.

In summary, we present the first experimental realization of switchable electric and magnetic skyrmions in THz TLPs. By employing nonlinear metasurfaces together with structured pump beams, we achieve active topological control, demonstrating both stability and tunability. Harnessing the synergy between TLP dynamics and the intrinsic robustness of skyrmions, this platform lays the foundation for next-generation communication systems, where information integrity is inherently safeguarded by topological principles. As topological photonics advances, our findings offer new prospects for quantum, and classical information communication.

**Data availability**

The data that support the findings of this study are presented in the paper and the Supplementary Information file.

**Acknowledgements**

This work is supported by the National Natural Science Foundation of China (Grant No. 62025504 (J.H.), 62135008(Q.X.), 62335011 (Q.X.) and 62405215 (X.C.) ); Singapore Ministry of Education (MOE) AcRF Tier 1 grants (RG157/23\&RT11/23 (Y.S.)); Singapore Agency for Science, Technology and Research (A*STAR) MTC Individual Research Grants (M24N7c0080 (Y.S.)); Nanyang Assistant Professorship Start Up Grant (Y.S.); China Postdoctoral Science Foundation (2024M752359 (X.C.)); Postdoctoral Fellowship Program of CPSF (GZC20241200 (X.C.)); Yunnan Expert Workstation (Grant No. 202205AF150008 (J.H.)).

**Conflict of interest**

The authors declare no competing interests.

**Contributions**

X.Z., Y.S. and J.H. conceived the idea. X.Z., L.N. and X.F. contributed to project conceptualization, methodology, and validation; L.N. and X.F. performed the measurements; L.N. and W.Y. carried out the theoretical calculations; L.N. and X.F. fabricated the samples; L.N. and X.Z. wrote the manuscript with suggestions from X.F., W.Y., Q.W., Y.L., Q.X., X.C, J.M., H.Q., Y.S., W.Z and J.H. X. Z. and J. H. supervised the project. All authors discussed the results and prepared the manuscript.

**Figure Captions**

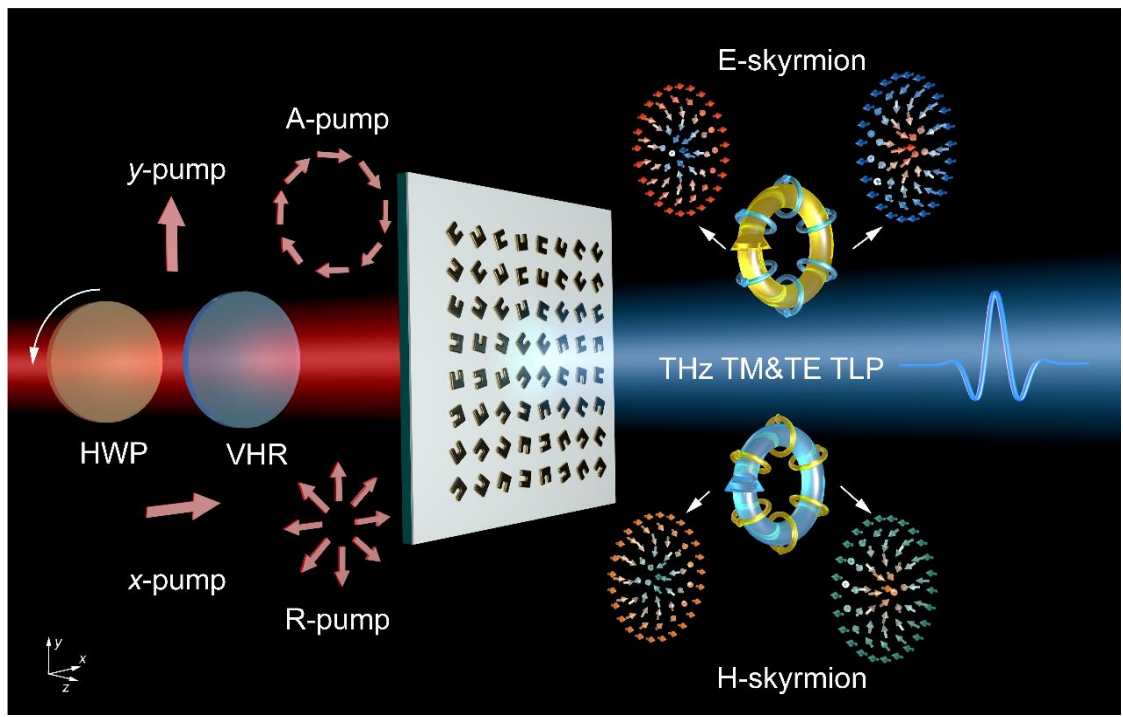

**Fig. 1 Schematic of the free-space E- and H-skyrmion switching in THz TLP generations via a nonlinear metasurface.** A *y*-pump is converted into an A-pump by a VHR, which excites the nonlinear metasurface to emit a TM TLP carrying a propagating E-skyrmion. By rotating the HWP, the incidence is switched to *x*-pump and converted into a R-pump by the VHR. This R-pump then drives the nonlinear metasurface to generate a TE TLP hosting a propagating H-skyrmion.

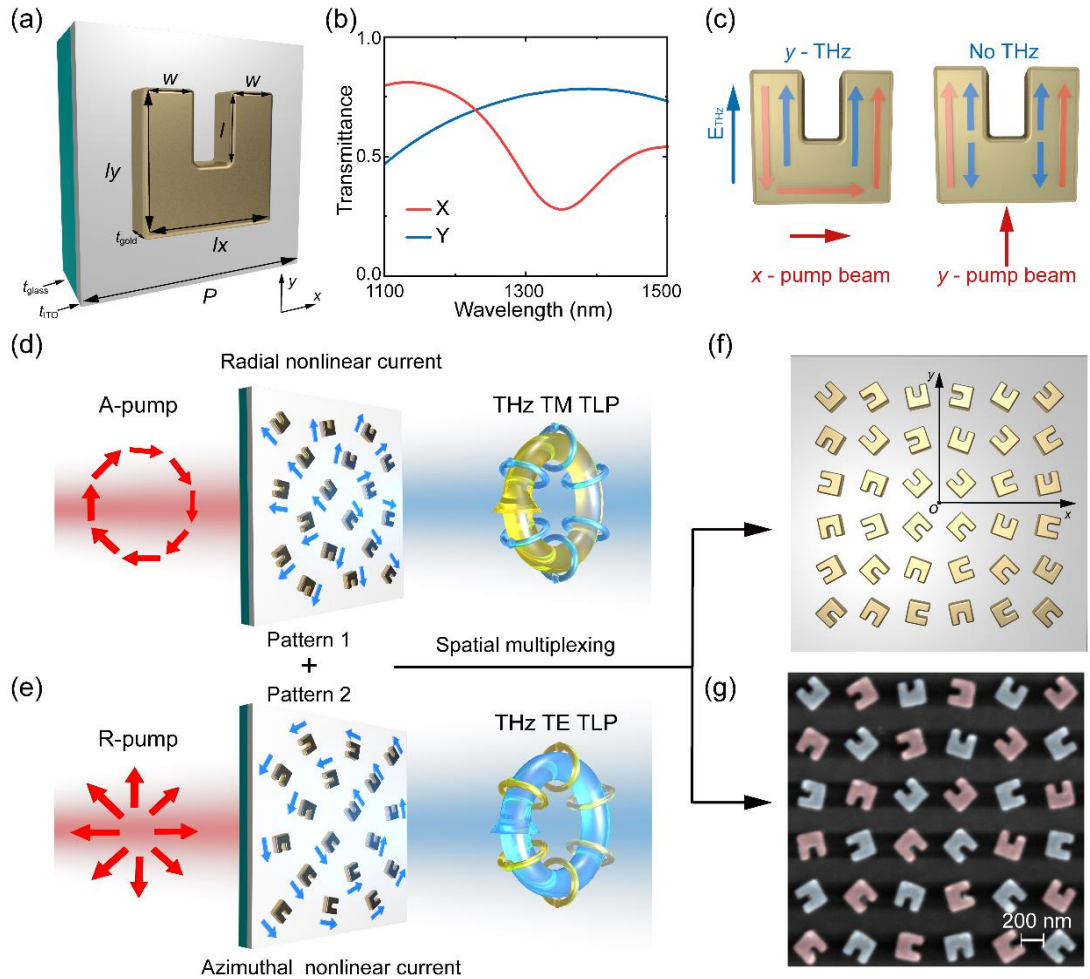

**Fig. 2 Design of the nonlinear metasurface. a** Schematic of the SRR unit cell, with dimensions of $l_x$ = 220 nm, $l_y$ = 212 nm, $l$ = 110 nm, $w$=79 nm, $t_{gold}$=40 nm, $t_{ITO}$ = 8 nm, $t_{glass}$ = 0.7 mm, and $P$ = 420 nm. **b** Simulated transmission spectra of the SRR under the *x*- and *y*-polarized illuminations. **c** Schematics of the linear and nonlinear current distributions under the *x*-pump and *y*-pump. **d** Under A-pump, the nonlinear metasurface with SRRs arranged in Pattern 1 generates a radially oriented nonlinear current, radiating a TM TLP. **e** Under R-pumping, the nonlinear metasurface with SRRs arranged in Pattern 2 generates an azimuthally oriented nonlinear current, radiating a TE TLP. **f** The final metasurface configuration achieved by spatially multiplexing the SRR arrangements of Pattern 1 and Pattern 2. **g** SEM image of the central region of the fabricated sample.

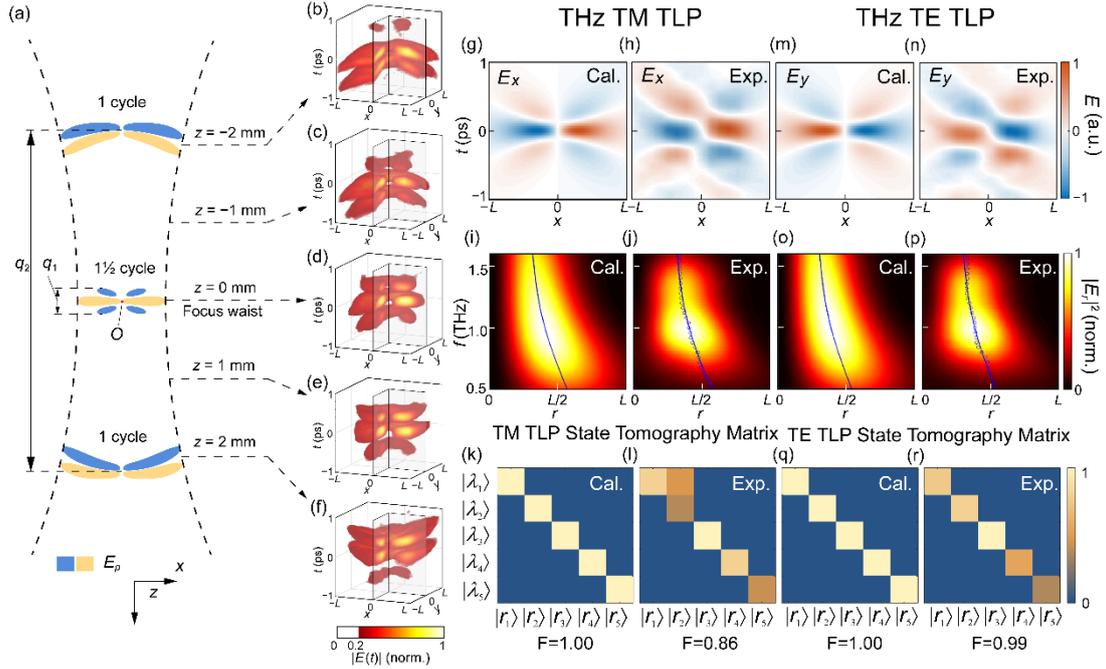

**Fig. 3 Characterization of the TM and TE THz TLPs emitted from the nonlinear metasurface. a** Schematic of the spatiotemporal distribution of the transverse electric field of the second-type THz TLP propagating along the $+z$ direction. **b-f** Measured normalized 3D spatiotemporal distribution of the THz transverse electric field amplitude at different $z$ positions. **g,m** Calculated and **h,n** measured normalized 2D spatiotemporal distribution of the THz transverse electric fields at $z = y = 0$. **i,o** Calculated and **j,p** measured normalized 2D space-spectral distributions of the THz intensity at $z = y = 0$. **k,q** State tomography of the ideal TM and TE toroidal pulses. **l,r** State tomography of the experimentally generated TM and TE THz TLPs. For all 3D maps, the electric fields in the region where $x > 0$ & $y > 0$ and $|E| = (|E_x|^2 + |E_y|^2)^{1/2} < 0.2$ are set to transparent to provide a clearer view of the internal field distribution.

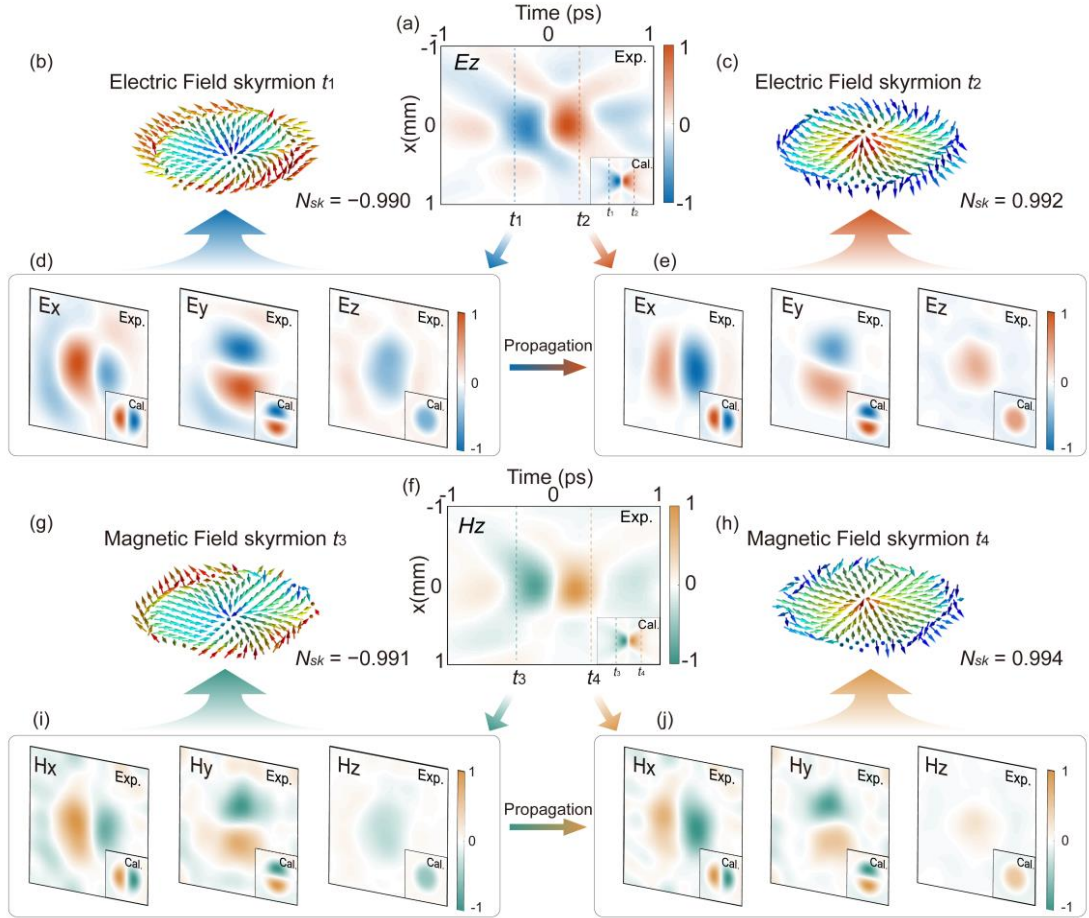

**Fig. 4 Measured results of the switchable E-skyrmion and H-skyrmion**. **a** Spatiotemporal map of the longitudinal electric field $E_z$ for the experimentally generated THz TM TLP at $z = y = 0$. **b,c** E-skyrmionic textures with preserved Néel-type helicity at different transverse planes at propagation times $t_1$ and $t_2$. **d,e** Distributions of $E_x$, $E_y$, and $E_z$ in the $xy$-plane at $t_1$ and $t_2$, respectively. **f** Spatiotemporal map of the longitudinal magnetic field $H_z$ for the experimentally generated THz TE TLP. **g,h** H-skyrmionic textures with preserved Néel-type helicity at different transverse planes at propagation times $t_3$ and $t_4$. **i, j** Distributions of $H_x$, $H_y$, and $H_z$ in the $xy$-plane at $t_3$ and $t_4$, respectively.